**Fluctuating ecological networks: a synthesis of maximum-entropy approaches for pattern detection and process inference**

Authors: **Tancredi Caruso[1*], Giulio Virginio Clemente[2] , Matthias C Rillig[3,4], Diego Garlaschelli[2,5]**

[1] School of Biology & Environmental Science, University College Dublin, Belfield, Dublin 4, Ireland

[2] IMT School for Advanced Studies, P.zza S. Francesco 19, 55100 Lucca, Italy

[3] Freie Universität Berlin, Institut für Biologie, 14195 Berlin, Germany

[4] Berlin-Brandenburg Institute of Advanced Biodiversity Research (BBIB), 14195 Berlin, Germany

[5] Lorentz Institute for Theoretical Physics, University of Leiden, The Netherlands

*corresponding author: tancredi.caruso@ucd.ie

**Authors' contribution statement**

TC and DG developed the concept of the paper, analysed the data and equally contributed to writing. GVC developed the Python scripts and contributed significantly to network analysis. MR critically reviewed the concepts and results of the work, and substantially contributed to writing.

**Abstract**

Ecological networks such as plant-pollinator systems and food webs vary in space and time. This variability includes fluctuations in global network properties such as total number and intensity of interactions but also in the local properties of individual nodes such as the number and intensity of species-level interactions. Fluctuations of species properties can significantly affect higher-order network features, e.g. robustness and nestedness. Local fluctuations should therefore be controlled for in applications that rely on null models, especially pattern and perturbation detection. By contrast, most randomization methods for null models used by ecologists treat node-level local properties as "hard" constraints that cannot fluctuate. Here, we synthesise a set of methods that resolves the limit of hard constraints and is based on statistical mechanics. We illustrate the methods with some practical examples making available open source computer codes. We clarify how this approach can be used by experimental ecologists to detect non-random network patterns with null models that not only rewire but also redistribute interaction strengths by allowing fluctuations in the null model constraints ("soft" constraints). Null modelling of species heterogeneity through local fluctuations around typical topological and quantitative constraints offers a statistically robust and expanded (e.g. quantitative null models) set of tools to understand the assembly and resilience of ecological networks.

**Introduction**

In recent years, there has been increasing recognition of the spatial and temporal variability of ecological networks, for example in terms of network rewiring (i.e. changes in who is connected to whom) in response to seasonality and perturbations (CaraDonna et al., 2017; Evans et al., 2013). At the same time, there are certain network features that remain rather stable. For example, plant-animal mutualistic networks can be highly variable in terms of which species is connected to which other species (i.e. their topology), while maintaining a clear core-periphery structure where a small number of generalist species, the core, interact with a large number of specialised species, the periphery (Miele et al., 2020).

The question of how ecological networks are assembled and what controls their stability has been intensely investigated (Aizen et al., 2016; Allesina & Pascual, 2008; Evans et al., 2013; Fortuna et al., 2010; James et al., 2012; Säterberg et al., 2013; Valdovinos, 2019; Valverde et al., 2020), typically by comparing network models to observed networks. Network models are typically assembled from data that describe who is interacting with whom (adjacency matrix) and possibly the intensity of interactions, i.e. link weights (interaction or weight matrix). The modelling usually starts from the observed interaction matrix with a focus either on structural patterns such as nestedness or the abundance of specific subgraphs/motifs, or the simulation of changes such as those due to species extinctions. In all these cases, changes in network structure are very often assessed with null models (Evans et al., 2013; Pascual & Dunne, 2006; Valdovinos, 2019).

Tailored null models based on constrained permutations of the values observed in the data matrix have become central to hypothesis testing and pattern detection in studies of ecological networks. One central goal is testing whether basic properties, for example at the node level (see Glossary), can explain higher-order properties such as nestedness (Bruno et al., 2020; Dormann et al., 2009; Dormann & Strauss, 2014; Payrató-Borras et al., 2019; Strona et al., 2014). Ecological null models generally permute, randomize, or sample the entries of network matrices by keeping certain values fixed, also known as constraints. Constraints can for instance be total sums over the entire matrix (e.g. total interaction strengths, or total number of links), or local sums along each row and/or column (e.g. the number of links of each node, known as the *degree*). The constraints are measured on the empirical matrix and generally enforced strictly, i.e. treated as "hard" constraints, in the construction of the random null model.

Here, however, we illustrate a different approach and propose that the intrinsic variability of ecological networks can be best modelled by allowing for the constraints to fluctuate, i.e. be assigned values that vary around the measured ones, and with the measured values representing a characteristic value (e.g. a mean value) for each node (species) in the network. The reason why we put emphasis on fluctuating constraints is twofold. First, the intrinsic variability of species-level activity, diet or behaviour implies that the interactions observed in a specific moment or experiment are a particular snapshot of a larger set of possible realizations, which is very evident in time series of the same system (e.g. Miele et al., 2020). For example, two species can be linked in one snapshot of the network and not linked in another. Or, a species might have a high number of connections in one snapshot but not in another snapshot. Also, experimental observations are necessarily subject to measurement errors such as spurious associations and missing data. Whatever the source of uncertainty, a cautious approach should interpret the measured entries of ecological matrices as particular realizations from a set of possible ones. Consequently, the constraints used in ecological null models should themselves fluctuate around a characteristic value. This perspective, while poorly explored in ecology (but see references we review in the next sections), is actually well established in application of statistical mechanics to the formulation of maximum-entropy ensembles of networks (Squartini & Garlaschelli, 2017). In the last two decades, ensembles of networks with fluctuating constraints have been developed theoretically in the general context of the statistical physics of complex networks (Cimini et al., 2019; Newman, 2018; Park & Newman, 2004), with applications mainly to large social, economic and financial networks. Some seminal applications to ecological networks have been recently proposed in the specific context of the property of nestedness (Bruno et al., 2020; Payrató-Borras et al., 2019; Payrató-Borràs et al., 2020).

Here, we propose that a general framework for the analysis of ecological systems both in terms of binary and weighted links is now possible and this paper aims to introduce and illustrate the existing theory and show applications of the existing models to ecological datasets. We emphasize that the approach we review here has important practical consequences for empirical analyses in ecology. For certain, simple systems, treating null constraints as fixed (or "hard"), as done in most ecological null models, or fluctuating ("soft") makes little difference when constraints represent *global* quantities such as the total number of links and when the matrix is very large. This property is known as *ensemble equivalence* (Touchette, 2015). However, a number of recent works have shown that ensemble equivalence does not hold when constraints are *local* (i.e. node or species specific), which is often the case of interest in modern applications and models, including for ecological networks which are typically very heterogenous at the level of local properties. This heterogeneity typically coexists, and is often responsible for other forms of hierarchy and structure such as core-periphery structure, the presence of network communities, or other types of modular organization (Garlaschelli et al., 2016; Squartini et al., 2015b; Zhang & Garlaschelli, 2020).

Why is network heterogeneity and the violation of *ensemble equivalence* important to the experimentalist? When networks are heterogeneous, statistical analyses based on models with hard local constraints (e.g. row and/or column sums that are kept exactly fixed), as mostly implemented by ecologists, can lead to different and even opposite results from those based on fluctuating constraints. This has been shown by a recent analysis of a large dataset of mutualistic plant-pollinator networks (Bruno *et al.* 2020). We also show this difference here with some practical examples and comparisons.

In this paper we illustrate the construction of network statistical ensembles with fluctuating constraints by presenting the general theoretical framework as well as some examples of application to real networks. The approach suits all types of ecological networks very well as it captures both their fundamental variability and experimental uncertainties associated with the measurement of the null model constraints themselves. Moreover, it is also computationally much more efficient than other currently available approaches, which is becoming increasingly important for molecular datasets, such as those linking microbes to plants. Also, and differently from current ecological null models, the approach we propose offers unbiased null models not only for presence/absence network data but also for networks with weighted links, which is a very important, new development currently underdeveloped in the ecologist toolkit for network. The possibility of generating unbiased null models for networks with weighted links is of major significance to systems such as food webs, where the links typically have heterogeneous intensities. Finally, we also propose several future research lines that we argue are needed to shed light on key questions on the processes that structure and control ecological networks

**Maximum-entropy ensembles of networks vs. rewiring algorithms**

The starting point of the approach we illustrate here is that the observed network matrix (which encodes both the topology and the link weights) represents the typical but not unique network state and that the network is best described by a large ensemble of possible states (Figure 1 and 2) fluctuating around the typical state. One state is a particular matrix realization and is assigned a probability of occurrence but how to derive this probability? The first step is the choice of the constraints. Constraints apply to the local properties of the nodes and not just to aggregated properties such as the total number of links in the network. That is because real-world networks neither are homogeneous (Caldarelli, 2007) nor symmetric upon arbitrary permutations of species (Miele et al., 2020). Some of the signatures of heterogeneity include the intrinsically hierarchical structure of ecological networks, e.g. the trophic hierarchy and allometric scaling of food webs (Garlaschelli et al., 2003), and the fact that most networks have broad distributions of the number of links per node (Bascompte, 2010).

In classical ecological null models, a typical constraint is the number of partners of a species, species by species, also known as degree sequence. In these models, the randomised matrices must respect the constraint exactly: if a species has, say, 10 partners in the observed matrix, it will have exactly 10 partners in all randomised matrices. These types of constraints are therefore "hard". After defining the hard constraints to be enforced, classical ecological null models generate an arbitrarily large set of random matrices computationally through a permutation rule or randomization algorithm (Camacho et al., 2007; Dormann et al., 2009; Gotelli, 2000). These techniques are also qualified as "rewiring algorithms" as they are fundamentally based on repositioning quantities within the original matrix, meaning that node to node links are rewired. There are, however, two criteria that must be respected by the null model: first, the randomization must ensure that locally exact constraints are met by the randomly rewired matrices (e.g. numbers of pollinator species associated with each plant species) and, second, the generated set of random matrices must be an unbiased sample of the probability distribution implied by the model, which typically has a non-closed form. Ensuring an unbiased sampling of the set of matrices alongside computational efficiency becomes a challenging algorithmic and combinatorics problem for increasingly large networks (Squartini et al., 2015a). Rewiring algorithms have in the past been used to generate such ensembles computationally including applications to food webs (Camacho et al., 2007; Stouffer et al., 2007). There are also important and statistically robust approaches developed for binary matrices (Carstens, 2015; Strona et al., 2014, 2018; Ulrich & Gotelli, 2012) allowing unbiased sampling of the random matrices set. Computational efficiency if not a major issue for small networks, especially if binary, but can become a major issue for unbiased sampling in the case of large networks. Even more fundamentally, samples from randomly rewired matrices have been shown to be statistically biased for heterogenous networks (Artzy-Randrup & Stone, 2005; Roberts &

Coolen, 2012) unless particular types of data structures and randomization algorithms are considered (Carstens, 2015; Strona et al., 2014, 2018; Ulrich & Gotelli, 2012). This means that local randomization algorithms carry a risk of not sampling the ensemble uniformly. The implication is that any quantity averaged over randomizations of the network (e.g. indices of nestedness in mutualistic network) is not guaranteed to correspond to the correct theoretical expectation of that quantity in the null model. Solutions to the issues of local rewiring algorithms are computationally intensive and apply only to specific conditions (Roberts & Coolen, 2012). Potentially, all these issues could be resolved in the future as computational power increases and more efficient algorithms are developed, which definitely applies to some types of not particularly large (i.e. rarely over a few hundreds of nodes) but very common ecological networks (Carstens, 2015; Strona et al., 2014, 2018; Ulrich & Gotelli, 2012). But the size of ecological datasets is becoming increasingly large with molecular sequence datasets. Also, the unequivocally robust randomization models apply only to topology (binary data) and cannot be naturally generalised to networks with link weights, or with a combination of purely topological and weighted constraints (Squartini et al., 2015a, Squartini & Garlaschelli, 2017). Another very important consideration, already mentioned earlier, is the fact that ensembles with soft local constraints are not equivalent to ensembles with hard local constraints (Bruno *et al.* 2020), which implies that even unbiased and efficient algorithms dealing with hard constraints are not a substitute for ensembles with soft constraints. This means that the choice of the soft vs. hard constraints is a fundamental modelling choice that will affect the final results of the analysis, as we show in our examples.

*The statistical mechanics solution*

A general solution to all these issues of null model formulation is offered by statistical mechanics (Cimini et al., 2019; Park & Newman, 2004). The key quantity to create a statistical mechanics ensemble is entropy. Entropy quantifies the uncertainty encoded in a probability distribution (see Glossary). The best-known expressions of entropy are Shannon entropy and Renyi's generalization through the so-called Hill numbers. In ecology, these expressions have been applied to various statistical distributions, e.g. in order to measure community diversity in terms of the balance between species richness and evenness (Hill, 1973; Magurran, 2013). There is also the maximum-entropy theory of ecology to explain patterns such as relative species abundance in communities (Harte, 2011; Harte & Newman, 2014). In the context of networks and statistical mechanics, Shannon (or equivalently Gibbs) entropy is applied to the probability distribution of a whole graph in an ensemble of possible ones. The graph probability distribution that maximizes the entropy under certain constraints reflects maximal ignorance of all network properties but those used to set the constraints themselves. This maximisation of entropy corresponds to the construction of networks that are maximally random, apart from the imposed constraints.

There are two fundamentally different ways in which constraints can be applied to derive a statistical mechanics ensemble. One way is the microcanonical ensemble, which enforces the constraint exactly on each randomly generated matrix. The microcanonical ensemble corresponds to the enforcement of hard constraints as used in classical ecological null models. For hard constraints, the maximum-entropy probability is uniform over the compatible configurations and the maximized entropy reduces to Boltzmann's definition of entropy, i.e. it equals the logarithm of the number of allowed configurations. For example, if the constraint was the degree sequence as observed in a real-world network, all the networks in the ensemble will have exactly the same degree sequence as the observed one. For binary networks, the model is known as the *microcanonical* (or "hard") *configuration model*, and its entropy is the logarithm of the number of graphs with given degree sequence. Finding this number remains a combinatorically challenging enumeration problem (Squartini & Garlaschelli, 2017). The approach we present here, instead, is called the canonical ensemble. In the canonical ensemble, the constraint is respected by the ensemble only on average and the investigator is looking at a system that fluctuates around a set of "typical" configurations, which are collectively the most likely. Again, if the constraint is the degree sequence, the corresponding model is known as the *canonical* (or "soft") *configuration model*. Its entropy can be interpreted as the logarithm of the effective number of typical configurations. When ensemble equivalence does not hold, this entropy is significantly different from the entropy of the corresponding microcanonical ensemble (Squartini et al., 2015b). For example, plant A might have five known pollinators, that is a degree equat to 5. Individual networks in the canonical ensemble might have the same species A with three or six, or any other number of pollinators, but such that the ensemble average of the degree of species A is exactly five. Plant A will thus have a theoretical average degree (i.e. weighted by the probability of all possible matrices in the ensemble matrix) exactly equal to five. The canonical ensemble can easily be sampled unbiasedly (Squartini et al. 2015a). That is, if the ensemble is sampled numerically and the average plant A degree is estimated as an arithmetic average over independently sampled matrices (i.e. using their frequency), the sample average will converge to five as the number of sampled matrices increases.

**The approach proposed here: canonical ensembles to model ecological networks**
*Matrix formalization*

We now formalize quantitatively the above notions and definitions. Let's call **O** (observed) the matrix that describes the observed network with $S$ species or nodes. The ensemble we are looking for consists of a large number of matrices, call each of them **E**$_i$ ("ensemble" matrices)**.** Each **E**$_i$ has the same size, i.e. the same number $S$ of nodes, as **O**. Moreover, across the entire ensemble, we preserve the identity of all nodes by attaching a unique label $i$ to each of them, i.e. the corresponding networks or matrices are *labelled*. In general, the difference between each **E**$_i$ and **O** is in terms of who is connected to whom and/or the strength of these connections. The set of the matrices **E**$_i$ are all possible states of the network. Among them, only one is exactly **O.** What characterizes the ensemble is the probability *P*(**E**) over the entire set of matrices, and we want this probability to depend on some structural properties of **O**. In particular, we choose a set of constraining properties using their values as empirically observed in **O**. As we want each constraining property to apply locally to each species, our constraint has the form of a vector, i.e. a vector having at least as many elements as the number of species in **O.** We say "at least" because we may want to have multiple constraints for each species, e.g. the number of incoming links (in-degree) and separately the number of outgoing links (out-degree) for each node in a directed network such as a food web. We denote the vector of constraints by **C**. We denote the value of property **C** attained on a generic network **E**$_i$ as **C(E**$_i$**)**, and so the empirical value of the constraint is denoted as **C\*= C(O)**, where the star means the special value of the constraining property as measured in the observed network **O.** For example, if **O** represents an undirected network and **C** is the degree sequence, then **C\*** will be the degree sequence of matrix **O**, i.e. the list of empirical degrees of all species, which would look something like **C\***=[$k_1, k_2, ... k_S$], where $k_j$ is the degree (number of interacting species) of each species *j* (for all *j*=1,*S*)*.*

*Entropy Maximization*

The main objective of statistical mechanics is finding the probability distribution *P*(**E**) that fulfils the constraint by guarantying maximum randomness of all other properties. With this probability distribution, we can formulate a statistical expectation of all observable quantities. Mathematically, finding the distribution *P*(**E**) that maximizes the randomness, given the constraints **C\*** (plus the additional constraint that P*(***E***)* has to be normalized), requires a quantitative definition of the randomness (i.e. uncertainty) encoded in *P*(**E**) in the first place. The statistical mechanics definition is Shannon (or Gibbs) entropy, defined as

$$S(P) \equiv -\sum_{\mathbf{E}} P(\mathbf{E}) ln P(\mathbf{E}) \quad \text{eq. 1}$$

which is familiar to ecologists as a diversity index. It is obvious from eq. 1 that entropy would just count the number of allowed states if *P*(**E**) were uniform. This is the same reason why the Shannon diversity index equals just species richness when all species are equally frequent and so the community has maximum evenness. Maximizing *S*(*P*) given the constraints, requires a choice on how specifically matrices $\mathbf{E}_i$ "realize" the constraint **C***, which is based on the observed matrix **O**. The two options are the aforementioned microcanonical or hard constraint and canonical or soft constraint ensembles respectively. The microcanonical ensemble is discussed in detail in the Supporting Information (part A) and is directly related to the classic ecological null models.

In the *canonical ensemble* the constraints are enforced as ensemble averages, i.e. as

$$\langle \mathbf{C} \rangle = \sum_{\mathbf{E}} \mathbf{C}(\mathbf{E}) P(\mathbf{E}) = \mathbf{C}^*, \quad \text{eq.2}$$

where $\langle \mathbf{C} \rangle \equiv \sum_E \mathbf{C}(\mathbf{E}) P(\mathbf{E})$ denotes the ensemble average (expectation value) of the property **C** across all matrices **E**. Now one has to find the form of *P*(**E**) that maximizes the entropy *S* in eq.1 subject to the constraint in eq.2, plus the additional normalization condition $\sum_E P(E) = 1.$ The result is well known and is obtained with the method of Lagrange multipliers (Park & Newman, 2004; Squartini & Garlaschelli, 2017). The maximum-entropy probability distribution depends on a vector **θ** of Lagrange multipliers and takes the form

$$P(\mathbf{E}|\boldsymbol{\theta}) = \frac{e^{-H(E,\boldsymbol{\theta})}}{Z(\boldsymbol{\theta})}, \quad \text{eq. 3}$$

where

$$H(\mathbf{E}, \boldsymbol{\theta}) \equiv \sum_j \theta_j C_j = \boldsymbol{\theta} \mathbf{C}(\mathbf{E})$$

Is the so-called *Hamiltonian* of the network, and

$$Z(\boldsymbol{\theta}) = \sum_E e^{-H(E,\boldsymbol{\theta})}$$

is the normalization constant, also known as *partition function*. Statistical physicists recognize the form of $P(\mathbf{E}|\boldsymbol{\theta})$ in eq.3 as the Boltzmann-Gibbs distribution over the graphs in the ensemble, where the Hamiltonian generalizes the energy (Park & Newman 2004), while statisticians and social scientists recognize it as the probability for the so-called Exponential Random Graph (ERG) models (Wasserman & Faust, 1994). The exact details of the derivation leading to eq.3 can be found elsewhere (Squartini & Garlaschelli 2017 and technical references therein) but what matters here is that we have introduced the new vector $\boldsymbol{\theta}$, the Lagrange multipliers, whose values have to be estimated in order to calculate the probability of each matrix $\mathbf{E}_i$ in the ensemble. Details on this calculation and estimates of $\boldsymbol{\theta}$ from the data using maximum likelihood are given in the Supporting Information (part B).

**Null models and pattern detection**

*Using the ensemble as a null model*

If the canonical ensemble respects the critical assumption set by the constraints (Figure 3), we can compare the observed network to the canonical ensemble for properties that were not used as constraints (Figure 2, 4 and 5).

We use the logic of null models and calculate the *z*-score of a network metric $X(\boldsymbol{E})$ as:

$$z_X = \frac{X(\boldsymbol{O}) - \langle X \rangle}{\sigma(X)} = \frac{X^* - \langle X \rangle}{\sigma(X)}$$

Where (again) $\boldsymbol{O}$ is the observed matrix, $X^* = X(\boldsymbol{O})$ is the observed value of the metric $X(\boldsymbol{E})$, while $\langle X \rangle$ and $\sigma(X)$ are the expected value and standard deviation, calculated under the probability $P(\mathbf{E}|\boldsymbol{\theta}^*)$ given by the null model, respectively. In principle, $\langle X \rangle$ and $\sigma(X)$ can be calculated analytically from $P(\mathbf{E}|\boldsymbol{\theta}^*)$; however, depending on the choice of $X(\boldsymbol{E})$, it may be more convenient to compute them as a sample average and a sample standard deviation respectively, over a large set of random matrices sampled numerically from $P(\mathbf{E}|\boldsymbol{\theta}^*)$. In either case, the z-score is simply the number of standard deviations by which the value of the metric in the observed network differs from its ensemble average. A positive score means that "observed X > expected X". If the distribution of the metric

X in the null model is normal, then the probability of observing a difference beyond two standard deviations just by chance would be roughly 0.05. Alternatively, if the ensemble metric is not normally distributed, a p-value can be defined following standard ecological null models (Gotelli 2000).

*Expanding the ecologist toolkit*

The form of equation 3 is very general. For applications, the researcher has to find the form of the graph Hamiltonian enforcing the constraints, which the randomised matrices of the ensemble will have to respect only on average ("soft"). Two examples of a model derivation are given in the Supporting Information (part C and D). Using the exponential family of random graphs, graph Hamiltonians and corresponding partition functions (eq. 3) have been derived not only for the ensemble corresponding to the usual constraints currently implemented by classic ecological null models but also many more others (Table 1). Importantly, we are interested in methods that apply not only to binary matrices (representing the mere topology of the network) but also to weighted matrices (which include information about link intensities). In weighted networks, each node is characterized not only by its degree but also by its strength (total weight of all links attached to that node: see Glossary). Consequently, besides the degree sequence, one can define the strength sequence as the vector of all node strengths. Canonical models allow imposing constraints on the degree sequence, the strength sequence, and also combined degree and strength sequences (Squartini et al 2015a, Squartini & Garlaschelli 2017), thereby filling a major gap in the current ecologist toolkit, which is mostly based on binary models or highly debated quantitative models. Canonical ensemble models can be fit by maximum likelihood using either the entire observed matrix or just partial information available on the constraints (e.g. just the degree and/or strength sequence) as input. In fact, to fit these models, only partial information on node-level properties is sufficient (e.g. just the degree sequence), even when there is no complete knowledge of the topology and/or the link weight distribution. The existing models can easily be fit using publicly available and computationally efficient routines written in Python, MatLab and C (see https://meh.imtlucca.it/ and the Python computer codes in the Supporting Information, which were used to generate the network ensembles of figure 3 to 5).

The most important types of models of interest to ecologists can be categorised in directed and undirect binary models (degree sequence as constraint, other topological properties are randomised), directed and undirected weighted models (strength sequence as constraint, other weighted and structural properties, including the degree sequence, are free to vary), and directed or undirected models where both degree sequence(s) and strength sequence(s) are constrained. Finally, the bipartite version of all these models, important for mutualist and host-parasite networks, can also be implemented by simple reparameterizations of the monopartite models. This very rich set of statistically unbiased models with soft constraints enormously extends the null model toolkits of ecologists, especially the models that allow randomising both topology and link weights subject to constraints on both degree and strength sequence. The possibility of implementing unbiased weighted null models is particularly important for weighted mutualistic network and food webs, where there typically is information on link weights. Also, the theory is flexible and general enough to generate new, future model ecologists will want to implement to test specific hypotheses. The implementation of specific constraints equate to find out the form of the graph Hamiltonian (eq. 3) that corresponds to the chosen set of constraints. For the ecological experimentalist, the task eventually reduces to fit a routine that corresponds to the tested hypothesis. This is another important advantage of the statistical mechanics approach over *ad hoc* randomization algorithms.

**Examples**

As an example, we applied bipartite models to a publicly available database of a plant-pollinator network (https://orcid.org/0000-0002-3449-5748), which has been recently proposed and analysed in the context of core-periphery models (Miele et al., 2020). This dataset consists of 6 years of data with three sampling time points each year, and the network links are weighted. For the purpose of illustrating the general feature of network canonical ensembles, we analysed both binary and weighted versions of one particular matrix in the dataset. Alongside the analysed data matrix (Supporting information, "bicm_mat.csv" for the binary version, and 'bicm_matW.csv' for the full weighted version) out in a format that can easily be imported in Python and R, we provide the Python codes used to fit the model (Supporting Information: Python script "Fit_BiCM" to fit a binary configuration model to a bipartite network, and "Fit_WBiCM.py to fit a weighted, bipartite configuration model) and the R codes used for downstream analysis (Bipartite.R and an Bipartite_W.R), which can altogether fully reproduce the results we present here from Figure 3 to Figure 5.

We estimated the ensemble using either just the degree sequence as constraint (Fit_BiCM.py" applied to "bicm_mat.csv"") or both the degree and strength sequence (Fit_WBiCM.py applied to 'bicm_matW.csv' ), to demonstrate how weighted models can also be implemented. For the binary case, we computed a classic null model based on hard constraints using a very well-known swap algorithm, which keeps the row and column margins fixed and thus randomises just species identity (Gotelli 2000). We also computed two null model using two swap algorithms for weighted matrices that can implemented by the R function "nullmodel" in the package "vegan". Specifically, through "commsim", we used the method "swap_count", which can preserve node strength sequence but not degree sequence, and the method abuswap_r, which preserves the node degree sequence but not the strength sequence.

Both for the canonical ensemble and classic null models, we sampled 999 matrices, checked that the main assumptions on the constraints were fully met by the ensembles and calculated various network metrics to compare the central tendency of the ensemble to the value of the metrics calculated on the observed matrix. In the case of the canonical ensemble, as the constraint is "soft", the ensemble respected the constraint only on average while the ensemble created with the swap algorithm consisted of matrices that all respected the imposed constraint exactly. For the weighted case, while the canonical ensemble respected both the degree and strength sequences (soft constraints met on average, see cross in Figure 3a), the ensemble created with the swap algorithm could respect (exactly, as "hard constraints) only either the strength or the degree sequence (Figure 3b), as expected. We are not aware of rewiring algorithms that can respect both degree and strength sequences as "hard" constraints while providing computational efficiency and an unbiased unform sampling of matrices with those constraints.

Once the ensembles are constructed, the ecologist can estimate metrics on the observed and randomised ensemble matrices, and then compare the observed metrics to the distribution of the randomised ones. Just as an example, for the binary model we chose NODF (a metric of nestedness), functional complementary, niche overlap and the so-called motif 5 as defined in Dormann & Strauss (2014) and Simmons et al. (2019). The canonical ensemble provided an excellent estimate of functional complementary and the so-called motif 5 (Figure 4a). Instead niche overlap and NODF (a metric of nestedness) deviated significantly from the ensemble average. What is really important, however, is that opposite results were observed for the classic null model based on the swap algorithm (Figure 4b), which had functional complementary and motif-5 deviating from the null model to a very large extent while NODF and niche overlap well replicated by the central tendency of the randomised matrices. For the weighted case, we calculated the weighted version of the clustering coefficient (Figure 5). Again, we observed opposite results for the canonical ensemble (Figure 5a) and the two swap algorithms we implemented (Figure 5b and 5c): the observed clustering coefficients greatly diverged from the central tendency of the canonical ensemble distribution, which was not observed with both rewiring algorithms.

What model is to be chosen then? We argue that the uncertainty in measured node level properties, which is due to a combination of measurement error and natural fluctuations, is statistically well handled by the soft constraint approach: a relatively large sample (e.g. 1000) of the null model ensemble returns an unbiased estimate of network metrics. That is not always the case for hard constraints both because a minimal error on the constraint values would bias the sampling and, more generally, because unbiased sampling of the output of rewiring algorithms is not achievable or achievable only under very specific cases for heterogeneous networks. And, finally, with soft constraints both topology and link weights can be randomised in a statistically robust and unbiased way, which is not possible with hard constraints.

**Conclusions**

Our examples show the importance of the assumptions behind the choice of whether hard or soft constraints should be imposed in the construction of null models. These assumptions are crucial to hypothesis testing. If the null model ensemble replicates higher order properties of the observed network, the implication is that the lower level properties (degree sequence and strength sequence in our models) used as constraints are sufficient to reconstruct at least some higher level properties (e.g. nestedness, clustering coefficient). There is thus a link between null model formulation and network reconstruction, which primarily aims at reconstructing unobserved or unobservable network properties using some of its fundamental features. To clarify the difference between traditional null models and network reconstruction, assume the investigators acquire some knowledge on the processes that assemble the network or have hypotheses on these processes. More specifically, the investigators have hypotheses on how these processes structure some fundamental properties of the network, say the degree or strength sequence. For example, there might be information on the total energy flux entering and leaving a species, but not on all the details of how this flux is partitioned across the resources and consumers of that species. The investigators then measure the observable properties and use the measurements as constraints to reconstruct the network, taking into account all the uncertainty in the details. Now, the investigator can test the quality of their network reconstruction using other (unconstrained) properties to test their hypothesis with the null model approach.

As shown by our examples and by other examples in ecology (Bruno et al., 2020; Payrató-Borras et al., 2019; Payrató-Borràs et al., 2020) as well as in several other disciplines (Cimini et al., 2019; Squartini & Garlaschelli, 2017), careful choices of the constraints can lead to canonical ensembles that replicate various other network properties. If we interpret the canonical ensemble as a reconstruction of the network, the processes that control the shape of constraints also control the probability distribution of the network configuration and all the properties that depend on this configuration. Also, a perturbation that changes the processes and factors that control network structure may move the network away from the typical configuration, which can be detected (Squartini et al., 2013) because the reconstruction is sometimes successful at predicting non-constrained properties but sometimes it is not. When it is not, the comparison between the empirical evolving network and the canonical null model can reveal and quantify the ongoing departure of the system from its typical state.

The most novel tool we highlight in this paper is that multiple topological and weighted observed properties of a network can be used to constrain the construction of the null model ensemble. The properties chosen as constraints can be enforced one at a time or all at the same time, making the ensemble more or less tightly dependent on the original network. The choice of the constraints is hypothesis-driven; enforcing different properties one at a time or at the same time can reveal the relative roles of the factors that contribute to the formation of the network.

**Acknowledgements**: the authors were supported the Strategic Priority Support Mechanism (project Re-EcoNet) of the Earth Institute, University College Dublin. The authors are grateful to two anonymous reviewers for their constructive comments on an earlier draft.

**Figure Captions**

**Figure 1** Construction of a statistical mechanics ensemble of a network, from the observed network (top) to the final ensemble (bottom). Starting from the observed network, node-level properties such as the node degree (number of connections to the node) are enforced to construct the ensemble. The illustration is based on animal pollinators (light blue) and their plants (light green). The constraint is used to find the probability distribution *P(E)* that maximises the entropy *S*. In the canonical ensemble, the constraint is enforced only on average. In this case, the maximization of the log-likelihood function defined through *P(E)* is used to find the *P(E)* parameterization that also maximises entropy given the constraint. The log-likelihood of the observed network equals the ensemble entropy (but a minus sign), and unbiased sampling of the ensemble is possible through the parameterized *P(E)*. See Figure 2 for the features of this sample.

**Figure 2.** Many matrices can be sampled from the ensemble to create a null distribution (histogram) for any network property. For example, metrics such as NODF (a measure of nestedness) can be computed for each of the *n* matrices sampled from the ensemble and on the single matrix of the observed network (blue vertical line). A z-score can also be calculated.

**Figure 3** Node degrees and node strengths of the observed network are jointly used as constraints to derive the canonical ensemble of the network (panel a). If the ensemble is estimated correctly, by construction the theoretical expected value of both degree and strength sequences should be exactly equal to the observed value (computationally, this should happen for an average over a large number of sampled matrices). Here, the cross symbols correspond to an average over 999 matrices from the canonical ensemble of the 8th matrix of the general plant-pollinator dataset of (Miele *et al.* 2020), used here as an example. It is evident that these averages fall on the identity line, confirming that the ensemble respects the key assumptions. The coloured dots correspond to various

randomly chosen sampled matrices, showing that the ensemble built from soft constraint is a set of networks whose properties (coloured dots) fluctuate around a mean, "equilibrium" state (cross symbols) which inherits the observed node level-properties. In b) and c) the same representation of a) but respectively for an ensemble obtained either with the "swap_count" algorithm (1) or with the "abuswap_r" algorithm (2) available in the nullmodel function of "vegan" in R . These quantitative swap algorithms preserve either the strength or the degree sequence exactly but not both strength and degree sequence. We are not aware of rewiring algorithms that can respect both degree and strength constraints and that can be applied to heterogeneous networks guaranteeing an unbiased (uniform) sampling of the ensemble.

**Figure 4** Four binary network metrics calculated for the a) canonical ensemble of the binary version of the network of Figure 3, from Miele *et al.* (2020) and b) the classic null model ensemble obtained with the "swap" algorithm" of "nullmodel" in vegan, which preserves row and column margins and is traditionally considered a robust model to test for the pure effect of species composition. The most evident result is that the two ensembles return opposite output and we advocate for soft constraints if there is uncertainty on the experimental measurement of the constraint and/or if the constraint is known to fluctuate.

**Figure 5.** As figure 4, for the same network but with weighted links (see figure 3). In this case, we calculated the weighted version of the clustering coefficient for the two network layers (plants and pollinators). With soft constraints (canonical ensemble, panel a), the null model is rejected, meaning that combined strength and degree sequence alone cannot replicate the weighted clustering coefficient. The two tested quantitative swap algorithms are the same as in Figure 3 and would lead to the opposite conclusion (null model retained), but we also know from Figure 3 that these two ensembles are not reconstructing either the degree or weight sequence, which could bias the reconstruction

of the clustering coefficient.

**Table 1.** Summary of existing canonical models that can be fitted to an ecological network matrix, either using just the degree sequence (binary), the strength sequence (weighted), or both sequences as "soft" constraints. The models are compared to counterparts with "hard" constraints, in particular some illustrative (but not exhaustive) examples of the most popular rewiring algorithms used by ecologists.

| Network type | Soft constraints: Canonical Ensemble | Hard constraints: Rewiring algorithms in ecological null models |
|---|---|---|
| Undirected: | | |
| Binary | UBCM [1] | e.g.: swap sequential algorithms, fixed margins |
| Weighted | UWCM, CReM [1] [2] | e.g.: swap sequential algorithms, fixed |
| Binary + weighted | UECM, CReM [1] [2] | only partially developed (e.g. swap and shuffle algorithms) |
| Directed: | | |
| Binary | DBCM, BiCM [1] [2] [3] | implicit within algorithms for undirected matrices |
| Weighted | DWCM, CReM [1] [2] | implicit within algorithms for undirected matrices |
| Binary + Weighted | DECM, CReM [1] [2] | not developed |

[1] Squartini et al. 2015; [2] Parisi et al. 2020. [3] Saracco et al. 2015.

**Glossary**

*Entropy*: a fundamental concept in thermodynamics, probability theory, statistical physics and information theory, for which many definitions exist. In statistical physics, the definition of entropy quantifies the amount of uncertainty encoded in a probability distribution representing the possible microscopic states of the system. Looking for the probability distribution that maximizes the entropy, subject to certain constraints (see definition in this Glossary), is a key step in the formulation of null models of networks that are maximally random, apart from a set of properties that are enforced (i.e. the constraints).

*Canonical network ensemble*: a maximum-entropy ensemble in which the constraints (see definition in this Glossary) are respected only as average values. Such constraints are called "soft". This means that individual realizations of the system (e.g. network matrices) sampled from the ensemble will in general not meet the constraints exactly. However, as a result of entropy maximization under the enforced constraints, some configurations are much more probable than others and the probability distribution is centred around the "typical" ensemble configurations that do meet the constraints. The ensemble average of the constraints exactly equals the enforced values. The resulting (maximised) entropy equals Gibbs's, or equivalently Shannon's, definition and can be interpreted as the logarithm of the effective number of typical configurations. The ensemble can be conceptualised as a set of networks "fluctuating around" the typical configurations formalised through the constraints, i.e. around the microcanonical configurations (*cfr* microcanonical ensemble definition in this Glossary).

*Constraints*: a set of properties that a (null) model must obey. In the context of entropy maximization, the constraints are used to find the maximally random probability distribution, while keeping certain properties fixed. The first and unavoidable constraint is the normalization condition guaranteeing that the result of the maximization is indeed a probability distribution. The additional constraints represent properties that one wants to enforce, e.g. structural network features that one would like to be identical to those empirically observed in a real network, while leaving maximum randomness in all the other properties of the network. This is the basis for separating random and nonrandom patterns in network structure. Global constraints are network-wide properties, such as the total number of links in the matrix. Local constraints, instead, are defined at the node level, such as the number of links to a specific node or the entire degree sequence (see definition in this Glossary).

*Degree and degree sequence*: in a network, the degree of a node is the number of links reaching that node. The degree sequence is a vector that, node by node, lists the degrees of all nodes, where the position in the list corresponds to the label of the node. As an example, the degree sequence $\{1,10,7,\ldots\}$ (of length S) indicates that nodes 1, 2 and 3 (out of the S nodes in the network) have degree 1, 10 and 7 respectively. If the network is directed, two distinct ("out-" and "in-") degrees per node can be defined, as the number of out-going and in-coming links respectively. Correspondingly, the out-degree sequence and the in-degree sequence can be defined as two separate vectors.

*Microcanonical network ensemble*: a maximum-entropy ensemble in which the constraints are respected exactly by each individual realization of the system (e.g. by each allowed state of the network). As a result, the constraints are called "hard". The maximum-entropy probability distribution of the microcanonical ensemble is uniform over all configurations that realize the hard constraints. The resulting (maximized) entropy coincides with the classic definition by Boltzmann: it equals the logarithm of the number of allowed configurations. It is generally very hard to calculate this number. For example, if the constraint is the degree sequence (see definition in this Glossary) and if the value of such constraint is set equal to the value of the degree sequence observed in a real food web, then each of the networks sampled from the microcanonical ensemble will have exactly the same degree sequence of the observed one. As a comparison, the randomization schemes usually implemented by ecologists also use "hard" constraints, in the sense that the randomised matrices respect these constraints exactly. However, not all algorithms with hard constraints (especially in the case of weighted and/or local constraints) are guaranteed to actually sample the correct maximum-entropy (uniform) distribution, and one should be aware of the risk of bias.

*Strength and strength sequence*: in a network with weighted links, such as a food web where the links can be expressed in units of energy or C fluxes, the strength of a node is the sum of all the weights of the links connected to that node. If the network is directed, two separate ("out-" and "in-") node strengths can be defined for each node. Similar to the degree sequence (see definition in this Glossary), the strength sequence is a vector that lists, node by node, the strength of that node. For directed weighted networks, the out-strength sequence and the in-strength sequence can be defined as two separate vectors.

Figure 1

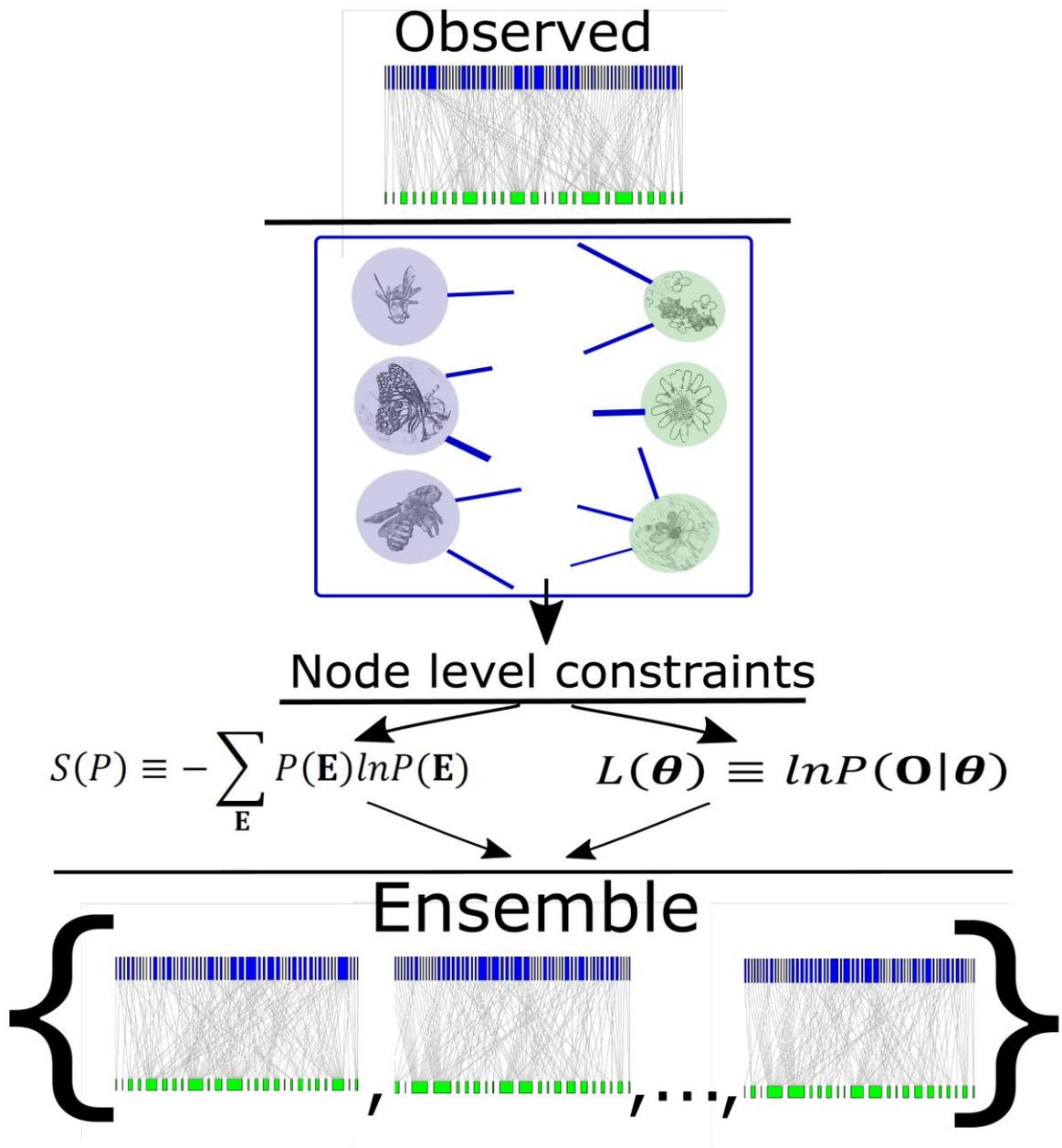

Figure 2

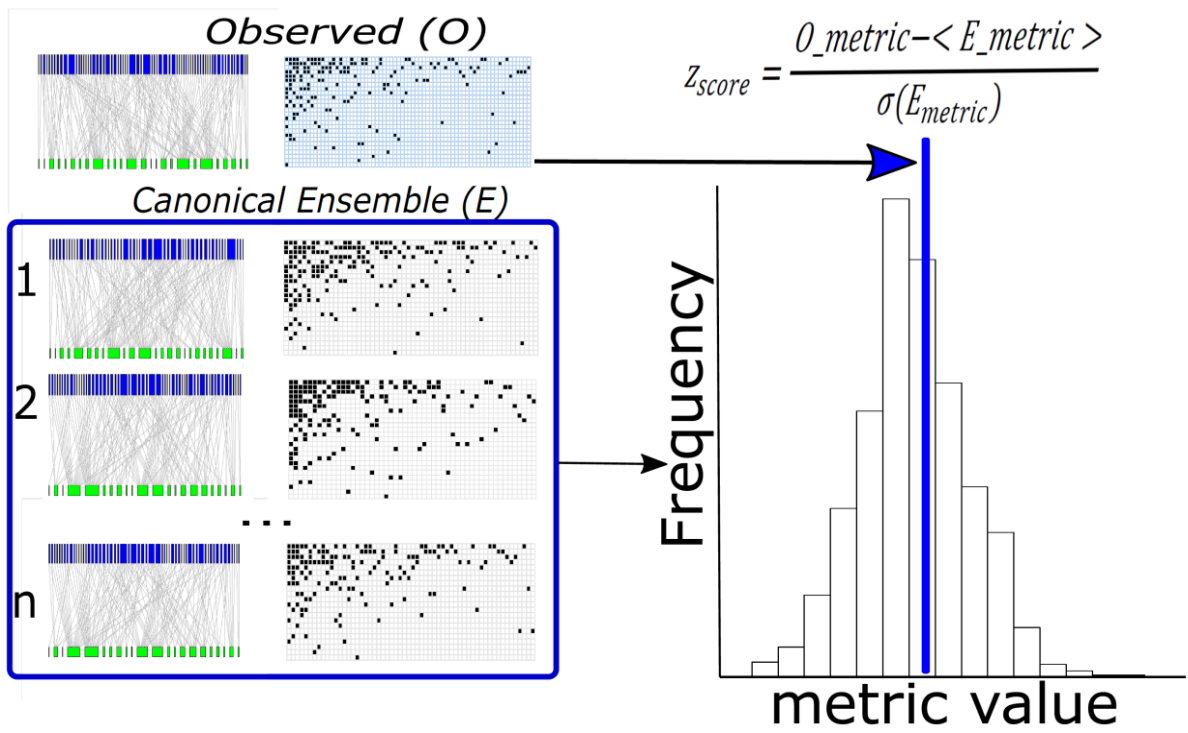

Figure 3

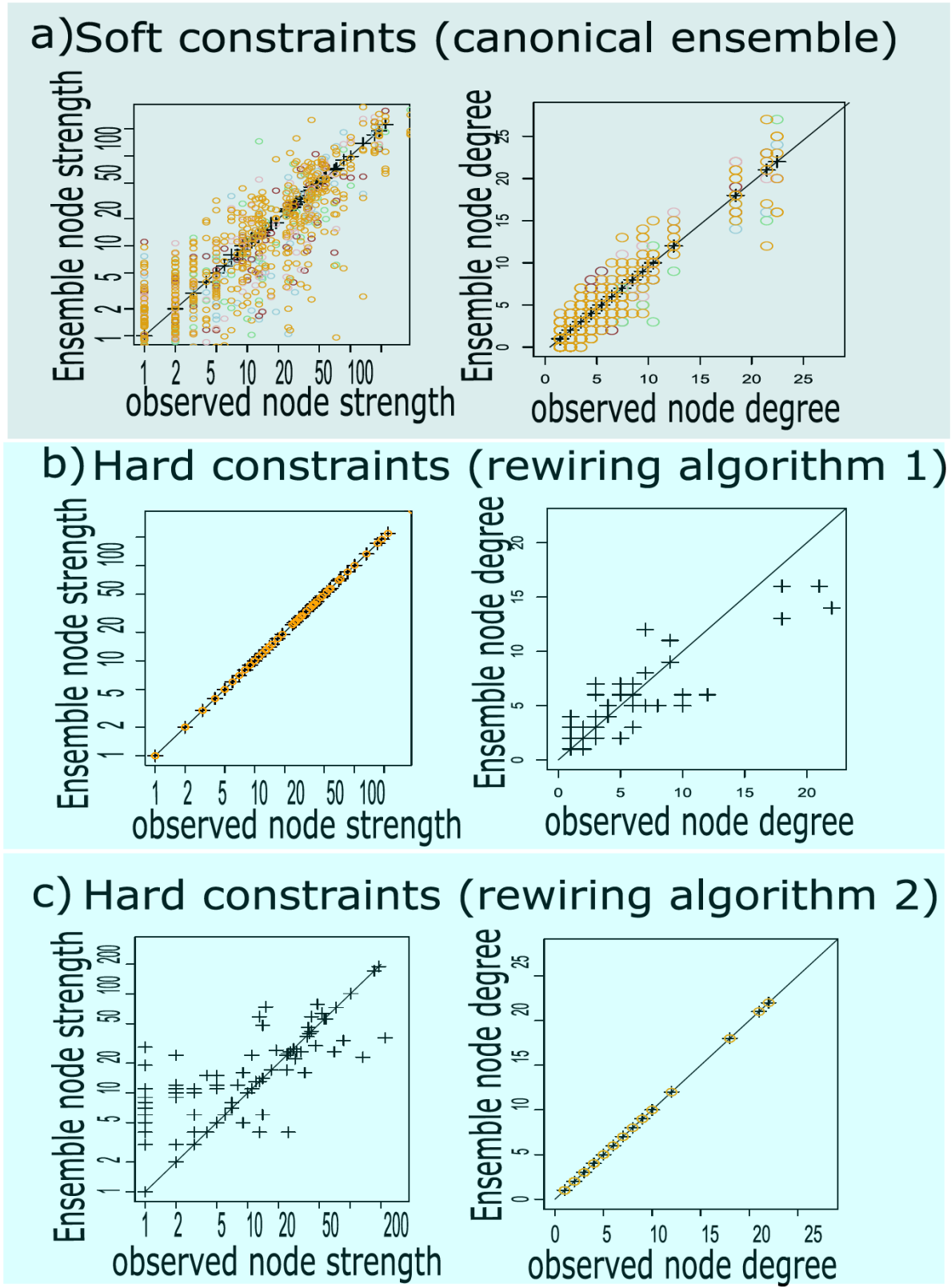

Figure 4

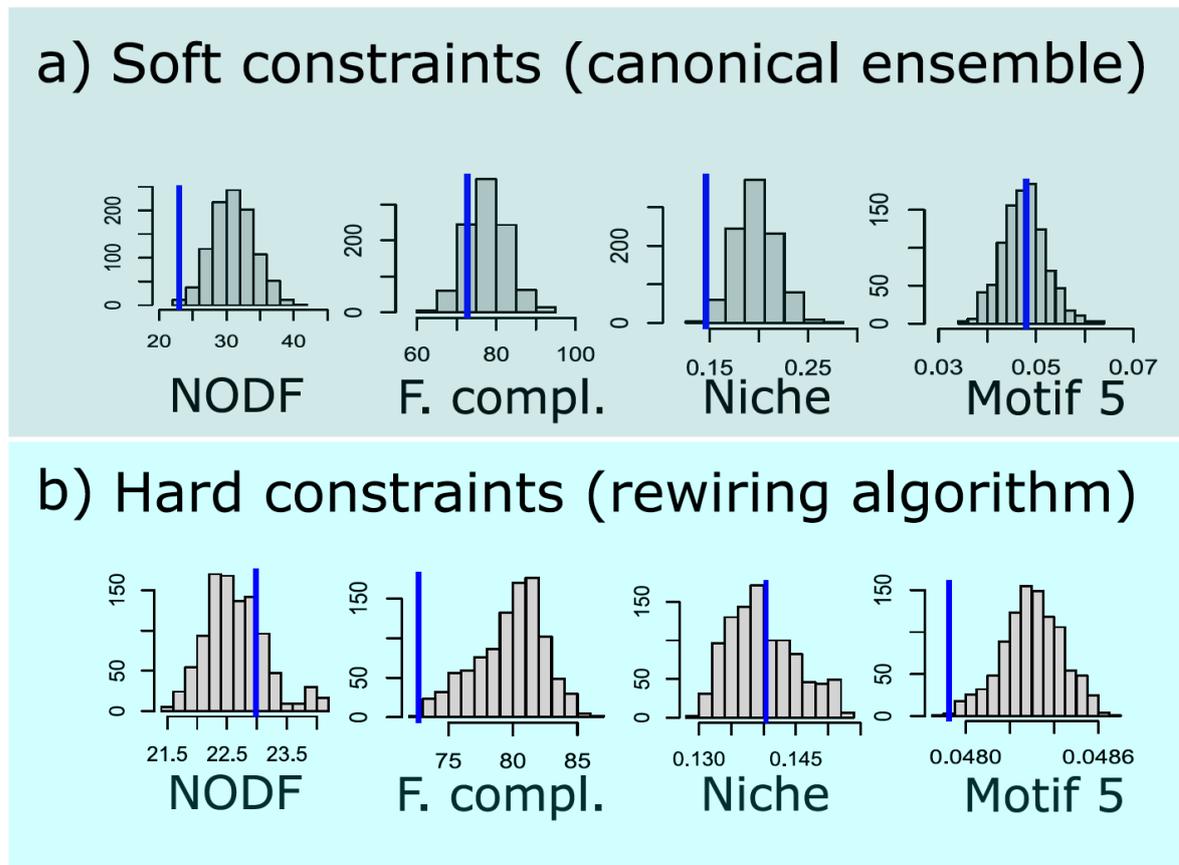

Figure 5

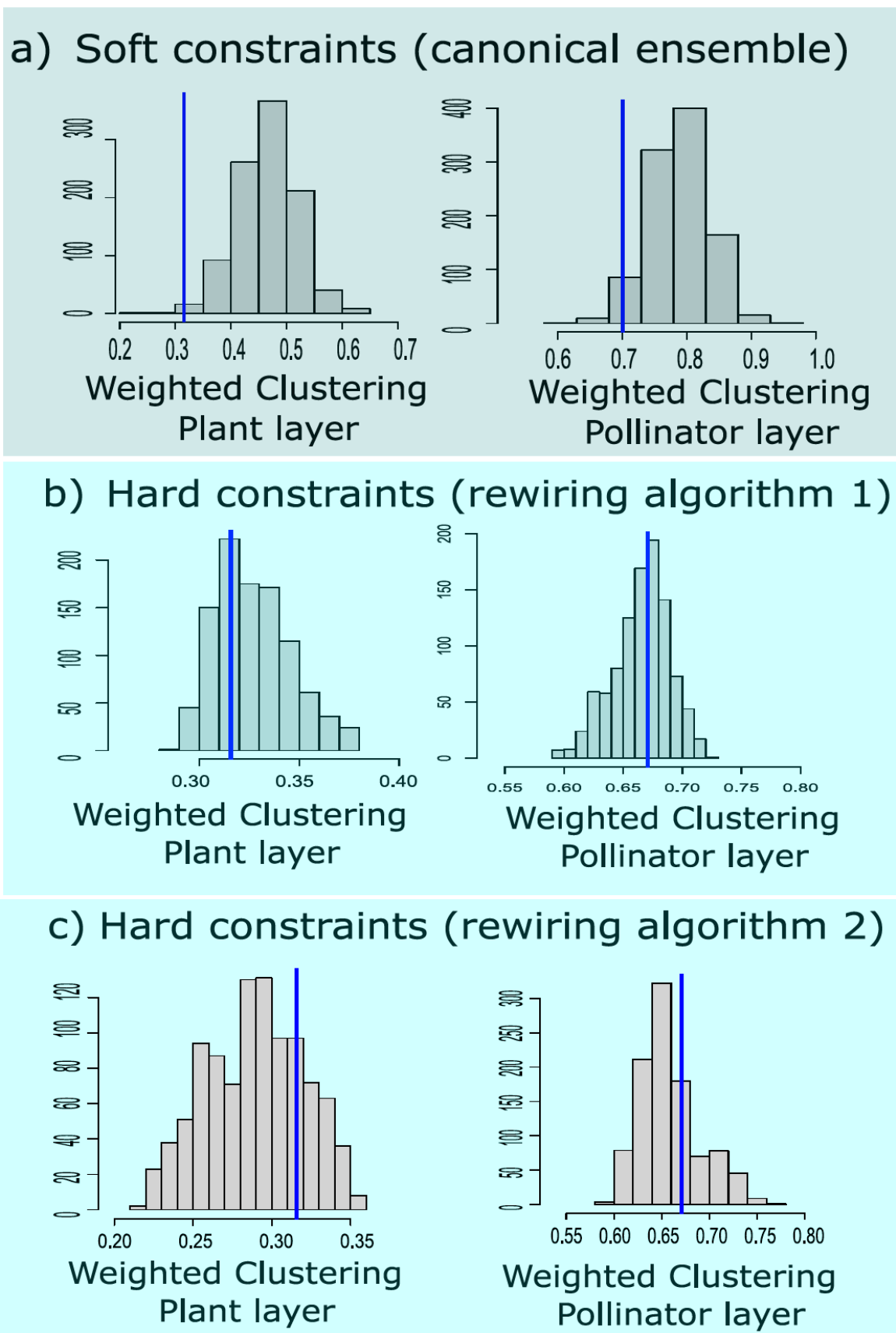